\documentclass[12pt]{article}
\usepackage{epsfig}

\usepackage{amssymb}
\usepackage{amsmath}
\usepackage{amsfonts}

\def\be{\begin{equation}}
\def\ee{\end{equation}}
\def\ba{\begin{array}{c}}
\def\ea{\end{array}}

\def\ben{$$}
\def\een{$$}

\newcommand{\bea}{\begin{eqnarray}}
\newcommand{\eea}{\end{eqnarray}}

\newcommand{\bbr}{\br\!\br}
\newcommand{\kkt}{\kt\!\kt}
\newcommand{\pbr}{\prec\!}
\newcommand{\ppbr}{\prec\!\prec\!}
\newcommand{\pkt}{\!\succ\,\,}
\newcommand{\ppkt}{\!\succ\!\succ\,\,}
\newcommand{\kt}{\rangle}
\newcommand{\br}{\langle}
\begin{document}

\title{${\cal PT}-$symmetric quantum systems with positive ${\cal P}$
%
}

\author{Miloslav Znojil\\
Nuclear Physics Institute ASCR, \\
250 68 \v{R}e\v{z}, \\
Czech
Republic\footnote{e-mail: znojil@ujf.cas.cz}\\
and
\\
Hendrik B.~Geyer\\
Institute of Theoretical Physics,
 University of Stellenbosch
and\\ Stellenbosch Institute for Advanced Study,\\ 7600
Stellenbosch, \\
South Africa\footnote{e-mail: hbg@sun.ac.za}}

\maketitle

\newpage

\section*{Abstract}

A new version of ${\cal PT}$-symmetric quantum theory is proposed
and illustrated by an $N$-site-lattice Legendre oscillator. The
essence of the innovation lies in the replacement of parity ${\cal
P}$ (serving as an indefinite metric in an auxiliary Krein space
${\cal K}^{\cal P}$) by its non-involutory alternative ${\cal
P}^{(positive)}:={\cal Q}>0$ playing the role of a positive-definite
nontrivial metric ${\cal Q}\neq I$ in an auxiliary, redundant,
unphysical Hilbert space ${\cal K}^{\cal Q}$. It is shown that the
${\cal P}^{(positive)}{\cal T}$-symmetry of this form remains
appealing and technically useful.

\vspace{5mm}

\noindent
{\it PACS:} 03.65.-w; 03.65.Ca; 03.65.Ta\\
{\it Keywords:} PT-symmetry; Hilbert-space metric; positive ${\cal
P}:={\cal Q}$; Legendre oscillator; quasi-Hermitian observables.

%

\newpage

\section{The formalism}

The operators of parity ${\cal P}$ and charge ${\cal C}$ entering
the product
 \be
 \Theta^{(PT)}={\cal P}{\cal C}
 \label{bend}
 \ee
play a key role in the ${\cal PT}-$symmetric quantum theory (PTQT)
of Bender {\it et al}  \cite{Carl}. The formalism may briefly be
characterized as a specific implementation of the standard quantum
theory in which the physical Hilbert space ${\cal H}^{(PT)}$ of
states is assumed Hamiltonian-dependent in the sense that the usual
``Dirac" (i.e., ``friendly but false" \cite{SIGMA}) inner product
$\br \eta|\Phi\kt^{(F)}$ of elements $|\eta\kt$ and $|\Phi\kt$ is
replaced by the Hamiltonian-adapted \cite{Geyer} inner product
 \be
 \br \eta|\Phi\kt^{(PT)}:=\br \eta|\Theta^{(PT)}|\Phi\kt^{(F)}\,
 \ee
which, under certain assumptions \cite{Sergii}, is {\em unique}.

In applications the systematically developed theory starts from an
arbitrary Hamiltonian $H$ with real spectrum which satisfies the
${\cal PT}-$symmetry requirement
 \be
 {\cal PT}H=H{\cal PT}\,
 \label{PTSym}
 \ee
where the anti-linear operator ${\cal T}$ mediates the reversal of
time \cite{BB}. The model is given its full physical
content by a self-consistent construction of a second observable,
viz., of the operator of charge ${\cal C}={\cal C}^{(PT)}(H)$
\cite{Geyer,BBJ}.

The chosen or ``input" Hamiltonian $H$ is allowed to be
non-Hermitian in the Dirac sense, $H \neq H^\dagger$. Bender and
Boettcher's \cite{BB} ${\cal PT}-$symmetric oscillator
$H^{(BB)}=p^2+gx^2$ with $g=g(x)=({\rm i}x)^\delta$ and $\delta>0$
serves as an often quoted example. Still, characterizing the PTQT
models primarily as non-Hermitian would be misleading. Indeed, these
models are strictly Hermitian in the physical Hilbert space ${\cal
H}^{(PT)}$ while only {\em appearing to be} non-Hermitian in the
(irrelevant and manifestly unphysical, {\em auxiliary})
representation space ${\cal H}^{(F)}$ endowed, by definition, with
the trivial, ``friendly but false" Dirac metric $\Theta^{(Dirac)}=I
\neq \Theta^{(PT)}$.

The overall methodical framework remains unchanged when we replace
the involutory operator of parity (with the property ${\cal P}^2=I$)
by another indefinite (and invertible) operator ${\cal P}$. In the
language of mathematics the correspondingly generalized ${\cal
PT}-$symmetry requirement (\ref{PTSym}) still remains practically
useful and mathematically tractable as equivalent to the so called
pseudo-hermiticity of~$H$ \cite{ali} or to the Hermiticity of $H$ in
another auxiliary space, viz., in a Krein space ${\cal K}^{\cal P}$
endowed with the metric  ${\cal P}$ (often called pseudo-metric)
which is, by definition, indefinite~\cite{Sergii,Langer}.

The parity-resembling pseudo-metric ${\cal P}$ may possess just a
finite number $k >0$ of anomalous (i.e., say, negative) eigenvalues.
Even then, the validity of Eq.~(\ref{PTSym}) entitles us to speak
about a PTQT model and, in a purely formal sense, about the
Hermiticity of~$H$ in the auxiliary Krein (or, at finite $k$,
Pontryagin) space ${\cal K}^{\cal P}$~\cite{cinani}.

In our present note we shall demonstrate specific methodical as well
as phenomenological appeal of the extreme (and, as it seems, not yet
contemplated) choice of the non-involutory invertible operator
${\cal P}$ which proves, in addition, positive definite (mimicking
the extreme choice of $k=0$). Of course, once we replace the symbol
${\cal P}$ (reserved, conventionally, for the indefinite $k>0$
pseudometrics) by ${\cal Q}$ (the symbol characterizing a positive
definite operator), our auxiliary space ${\cal K}^{\cal Q}$ will
cease to be a Krein (or Pontryagin) space, re-acquiring the
perceivably less complicated Hilbert-space mathematical status.

In the PTQT spirit, the {\em physical} status is expected to be
re-acquired by some {\em other}, {\it ad hoc} Hilbert space ${\cal
H}^{(PT)}$ (or rather ${\cal H}^{(QT)}$, to be specified below). In
other words, we shall describe here just a version of the PTQT
model-building scheme in which

\begin{itemize}

\item [[p1\!\!\!]]
the ``input" operator ${\cal P}^{(positive)}={\cal Q}$ will play the
role of a positive-definite metric determining an auxiliary, {\em
unphysical} Hilbert space  ${\cal K}^{\cal Q}$;

\item [[p2\!\!\!]]
the ``input" Hamiltonian $H$ will be assumed ${\cal QT}-$symmetric,
i.e., compatible with Eq.~(\ref{PTSym}) where ${\cal P}$ is replaced
by ${\cal Q}$;

\item [[p3\!\!\!]]
we shall follow Ref.~\cite{ali} and reinterpret Eq.~(\ref{bend}) as
a mere {definition} of the operator ${\cal C}^{(QT)}={\cal
Q}^{-1}\Theta^{(QT)}$, with a {\em true} physical metric
$\Theta^{(QT)}\neq {\cal Q}$ {yet to be specified};

\item [[p4\!\!\!]]
finally, we shall recall Ref.~\cite{Geyer} and broaden the class of
admissible observables yielding, typically, a weakening of the
physical role played by ${\cal C}^{(QT)}$ itself.

\end{itemize}

 \noindent

\section{An exactly solvable illustrative example \label{solvable}}

A solvable example with properties [p1] -[p4] is derived here from
the recurrences which are satisfied by the Legendre orthogonal
polynomials $P_n(x)$. These recurrences may formally be rewritten as
an infinite linear system $ {H}^{(\infty)}|\psi_n\kt =
E_n|\psi_n\kt$ satisfied by the column vector-like array
 \ben
 |\psi_n\kt=\left (
  \ba
 \left (|\psi_n\kt\right )_1\\
 \left (|\psi_n\kt\right )_2\\
 \left (|\psi_n\kt\right )_3\\
 \vdots
 \ea
 \right )=\left (
  \ba
 P^{}_0(E_n)\\
 P^{}_1(E_n)\\
 P^{}_2(E_n)\\
 \vdots
 \ea
 \right )=\left (
  \ba
 1\\
 E_n\\
 \frac{1}{2}(3E_n^2-1)\\
 \vdots
 \ea
 \right )
 \label{inputas}
 \een
at an arbitrary real parameter $E_n$ (the range of which is usually
restricted to the interval $(-1,1)$ \cite{AS}). The system's real
matrix
 \ben
 {H}^{(\infty)}=
  \left[ \begin {array}{cccccc}
   0&1&0&0&0&\ldots
  \\
  1/3
 &0&2/3&0&0&\ldots
 \\0&2/5&0&3/5&0&\ldots
 \\0&0&3/7&0&4/7&\ddots
 \\0&0&0&4/9&0&\ddots
 \\
 \vdots&\vdots
 &\ddots&\ddots&\ddots&\ddots
 \end {array} \right]\,
 \label{kitiel}
 \een
is asymmetric and, therefore, it cannot be interpreted as Hermitian
in the Dirac sense.

\subsection{Requirements [p1] and [p2]}

The infinite system of equations ${H}^{(\infty)}|\psi_n\kt =
E_n|\psi_n\kt$ is not suitable for our present purposes because all
of its formal ``eigenvectors" $|\psi_n\kt$ would have an infinite
Dirac norm. This can be deduced from the completeness of the
Legendre-polynomial basis in $L^2(-1,1)$. Thus, we have to turn
attention to a truncated, finite-dimensional version of our
linear system.

Firstly we find that using the $N$ by $N$ truncated submatrices
${H}^{(N)}$ of ${H}^{(\infty)}$ at any $N=1,2,\ldots$, the
corresponding Schr\"{o}dinger-like equation
 \be
 {H}^{(N)}|\psi_n^{(N)}\kt = E_n^{(N)}|\psi_n^{(N)}\kt\,,\ \ \ n = 0, 1,
 \ldots, N-1\,
 \label{sejed}
 \ee
may only be satisfied by the truncated vector (\ref{inputas}) if we
manage to guarantee that the next-neighbor vector element vanishes,
 \be
 \left (|\psi_n\kt\right )_{N+1}=P^{}_{N}(E_n)=0\,.
  \ee
{\it Vice versa}, the $N-$plet of roots of the latter polynomial
equation is nondegenerate and real so that it strictly coincides
with the spectrum of ${H}^{(N)}$. The eigenvalue part of the problem
is thus settled.

Secondly, once we eliminate ${\cal T}$ from Eq.~(\ref{PTSym}) and
replace ${\cal P}$ by  ${\cal Q}$ yielding
 \be
 \left [ {H}^{(N)}\right ]^\dagger{\cal Q}=
  {\cal Q} {H}^{(N)}
  \label{inkspace}
  \ee
we may conclude that all of our Hamiltonians ${H}^{(N)}$ are ${\cal
QT}-$symmetric (i.e., compatible with Eq.~(\ref{inkspace})),
provided only that the specific choice is made of the diagonal and
positive definite matrix
 $
{\cal Q}= {\cal Q}^{(N)}
 $
with the truncation-independent matrix elements
 \be
  \left[{\cal Q}\right ]_{11}=\frac{1}{2}
 \,,\ \
  \left[{\cal Q}\right ]_{22}=\frac{3}{2}
 \,,\ \ \ldots\,, \ \ \
  \left[{\cal Q}\right ]_{NN}=\frac{N-1/2}{(N-1)!}\,.
  \label{inkspaceria}
 \ee
At any fixed $N$ this matrix is positive definite and may be
interpreted as a metric in Hilbert space, therefore.

Once this metric is used to define the inner products, the resulting
Hilbert space will be treated here as a direct parallel of the
auxiliary Krein space ${\cal K}^{\cal P}$ of PTQT. This parallelism
will be underlined not only by the notation (our Hilbert space with
the inner products defined via matrices (\ref{inkspaceria}) will be
denoted by the symbol ${\cal K}^{\cal Q}$) but also by the
interpretation (preserving the analogy, also our auxiliary Hilbert
space ${\cal K}^{\cal Q}$ will be assumed and declared {\em
unphysical}).

\subsection{Requirements [p3] and [p4]}

We assume that the time evolution of our quantum system {\em in spe}
(more precisely, of all of its eligible finite-dimensional ket
vectors $|\psi^{(N)}\kt$) is generated by the Hamiltonian, the
manifest non-Hermiticity  ${H}^{(N)}\neq \left [ {H}^{(N)}\right
]^\dagger$ of which forces us to complement Eq.~(\ref{sejed}), for
the sake of mathematical completeness, by the second, isospectral
Sch\"{o}dinger equation
 \be
 \left [ {H}^{(N)}\right ]^\dagger
 |\psi_n^{(N)}\kkt = E_n^{(N)}|\psi_n^{(N)}\kkt\,,\ \ \ n = 0, 1,
 \ldots, N-1\,.
 \label{sedru}
 \ee
Its ``ketket" solutions  $|\psi_n^{(N)}\kkt$  are in general different
from the kets $|\psi_n^{(N)}\kt$ of Eq.~(\ref{sejed})
\cite{SIGMA}. Nevertheless, the ${\cal QT}-$symmetry requirement
(\ref{inkspace}) may be recalled to simplify the second
Schr\"{o}dinger Eq.~(\ref{sedru}),
 \be
  {H}^{(N)}
 \left [{\cal Q}^{-1}|\psi_n^{(N)}\kkt
  \right ] = E_n^{(N)}\left [{\cal Q}^{-1}|\psi_n^{(N)}\kkt
  \right ]\,,\ \ \ n = 0, 1,
 \ldots, N-1\,.
 \label{sedruver}
 \ee
Due to the non-degenerate nature of the spectrum, the bracketed
vectors $\left [{\cal Q}^{-1}|\psi_n^{(N)}\kkt  \right ]$ must be
proportional, at every subscript, to their predecessors given by the
first Schr\"{o}dinger Eq.~(\ref{sejed}).

We are free to choose the proportionality constants equal to one. In
other words, the second Schr\"{o}dinger equation (\ref{sedru}) may
be declared solved by ketkets
 \be
 |\psi_n^{(N)}\kkt = {\cal Q}|\psi_n^{(N)}\kt\,,\ \ \ n = 0, 1,
 \ldots, N-1\,.
 \label{source}
 \ee
This is a useful convention because now, the {\em complete} {set} of
the eligible PTQT (or, rather, QTQT) metrics may be defined by
Mostafazadeh's expression \cite{Ali,SIGMAdva}
 \be
 \Theta^{(QT)}\left (H^{(N)}\right )=
 \sum_{j=0}^{N-1}\,|\psi_j^{(N)}\kkt\
 \kappa_j^{(QT)}\,\bbr \psi_j^{(N)}|\,.
 \label{defini}
 \ee
As long as we work with the finite dimensions $N < \infty$, our
choice of the $N-$plet $\vec{\kappa}^{(QT)}$ of the {\em optional,
variable} parameters $\kappa_j^{(QT)}>0$ is entirely arbitrary.

\subsection{QTQT models with nontrivial ``charge" ${\cal C}^{(QT)}\neq I$}

In the next step we take the elementary
decomposition of the unit operator
 \be
 I^{(N)} =
 \sum_{j=0}^{N-1}\,|\psi_j^{(N)}\kt
 \frac{1}{\br \psi_j^{(N)}|{\cal Q}|\psi_j^{(N)}\kt}\,\br
  \psi_j^{(N)}|{\cal Q}\,
 \ee
and multiply it by ${\cal Q}$ from the left, yielding another
identity
 \be
 {\cal Q} ={\cal Q}^{(N)} =
 \sum_{j=0}^{N-1}\,|\psi_j^{(N)}\kkt
 \frac{1}{\br \psi_j^{(N)}|{\cal Q}|\psi_j^{(N)}\kt}\,\bbr
  \psi_j^{(N)}|\,.
 \ee
A comparison of this expression with Eq.~(\ref{defini}) specifies the
exceptional set of constants
 \be
 \kappa_j^{(exc.)}=\frac{1}{\br \psi_j^{(N)}|{\cal
 Q}|\psi_j^{(N)}\kt}\,,\ \ \ j = 0, 1,
 \ldots, N-1\,
 \ee
at which the charge becomes trivial,
 ${\cal C}^{(exc.)}= I$, and
at which our two alternative Hilbert spaces, viz., spaces ${\cal
K}^{\cal Q}$ and ${\cal H}^{(QT)}$ would coincide.

{\it Vice versa},  a nontrivial QTQT operator ${\cal C}^{(QT)}={\cal
Q}^{-1}\Theta^{(QT)}\neq I$  is obtained whenever we choose, in
(\ref{defini}), any other $N-$plet of parameters,
 \be
 \exists j\,,\ \ \
 \kappa_j^{({\cal QT})}\neq \kappa_j^{(exc.)}
 =\frac{1}{\br \psi_j^{(N)}|{\cal Q}|\psi_j^{(N)}\kt}\,,
 \ \ \ 1 \leq j \leq N-1\,.
 \ee
The equivalence between our two alternative Hilbert spaces ${\cal
K}^{\cal Q}$ and ${\cal H}^{(QT)}$ becomes broken. In parallel, the
nontrivial QTQT operator ${\cal C}^{(QT)}\neq I$ must be found an
appropriate interpretation (obviously, calling it still a ``charge"
could be misleading). Summarizing \cite{SIGMA}, we may now comply
with the overall PTQT or QTQT philosophy and declare just the latter
metric $\Theta^{(QT)}$ and the related standard Hilbert space ${\cal
H}^{(S)}$ ``physical".

%

\section{Discussion}

\subsection{The criterion of observability in  ${\cal H}^{(QT)}$}

One of the remarkable consequences of the most popular choice of the
metric $\Theta^{(Dirac)}=I$ is that it trivializes the test of the
Hermiticity of any given set of observables $\Lambda_j$ in the
``friendly" space ${\cal H}^{(F)}$. On the contrary, whenever we
select a nontrivial $\Theta^{(QT)}\neq I$ in the physical space
${\cal H}^{(QT)}$, a similar test requires the verification of
validity of the Dieudonn\'e's \cite{Dieudonne} equation(s)
 \be
 \Lambda_j^\dagger \Theta^{(QT)}=\Theta^{(QT)}\,\Lambda_j\,
 \label{Dieud}
 \ee
where $\Theta^{(QT)}$ is given by formula (\ref{defini}). Thus,
assuming that we know the (curly bra-ket denoted) eigenstates and
spectral representation of
 \be
 \Lambda =
 \sum_{j=0}^{N-1}\,|\lambda_j^{(N)}\pkt
 \frac{\lambda_j}{\pbr \lambda_j^{(N)}|\lambda_j^{(N)}\ppkt}\,\ppbr
  \lambda_j^{(N)}|\,
 \ee
we may introduce two overlap matrices $U(\vec{\kappa})$ and
$V(\vec{\lambda})$ with elements
 \be
 U_{jk}(\vec{\kappa})=\frac{1}{\kappa_j}
 \br \psi_j^{(N)}|\lambda_k^{(N)}\ppkt\,,\ \ \ \
 V_{jk}(\vec{\lambda})=\frac{\lambda_j}{\pbr \lambda_j^{(N)}|\lambda_j^{(N)}\ppkt}\,
 \pbr \lambda_j^{(N)}|\psi_k^{(N)}\kkt\,
 \ee
and rewrite Eq.~(\ref{Dieud}) as the requirement of the Hermiticity
of the product
 \be
 U(\vec{\kappa})V(\vec{\lambda})=M=M^\dagger\,.
 \label{crite}
 \ee
A simplified version of the observability criterion (\ref{crite})
may be developed for the special Hilbert spaces ${\cal H}^{(QT)}$ in
which the nontrivial physical metric $\Theta^{(QT)}\neq I$ acquires
a special sparse-matrix form. Incidentally, our toy-model
Hamiltonians $H^{(N)}$ of section \ref{solvable} prove suitable for
an explicit constructive illustration of such an anomalous scenario.

\subsection{The emergence of hidden horizons}

A straightforward application of symbolic-manipulation software
reveals and confirms that at any dimension $N$, the
tridiagonal-matrix restriction imposed upon Eq.~(\ref{defini})
yields the one-parametric family of the tridiagonal
Legendre-oscillator metrics
 \be
 \Theta^{(QT)}_{(sparse)}={\cal Q}_{(diagonal)}+
 \left( \begin {array}{cccccc}
 0&\kappa_1&0&\ldots&0&0\\
 \kappa_1&0&\kappa_2&0&\ldots&0\\
 0&\kappa_2&0&\kappa_3&\ddots&\vdots\\
 \vdots&\ddots&\ddots&\ddots&\ddots&0\\
 0&\ldots&0&\kappa_{N-2}&0&\kappa_{N-1}\\
 0&0&\ldots&0&\kappa_{N-1}&0
 \end {array} \right)\,.
 \label{tridia}
 \ee
where $\kappa_j=\alpha/(j-1)!$ while ${\cal Q}_{(diagonal)}$ is
defined by Eq.~(\ref{inkspaceria}). In such a particular
illustration the condition of positivity of metrics (\ref{tridia})
restricts the admissible range of the optional parameter to a finite
and $N-$dependent interval, $\alpha\in (-\gamma^{(N)},
\gamma^{(N)})$. The boundary values $\gamma^{(N)}$ must be
determined numerically. In our tests they even seemed to converge to
a positive limit $\gamma^{(\infty)}>0$.

From the point of view of physics the latter observation (which
follows from our special choice of the form of the metric
(\ref{tridia})) might look like a paradox. Indeed, we know that the
spectrum of energies themselves remains real at any $\alpha$.
Fortunately, an explanation of such an apparent paradox is simple.
Once we have fixed the form of the metric
$\Theta^{(QT)}_{(sparse)}=\Theta^{(QT)}_{(sparse)}(\alpha)$ which
ceases to be invertible at $\alpha = \pm\gamma^{(N)}$, we may simply
connect this irregularity, via Dieudonn\'e's Eq.~(\ref{Dieud}), with
the loss of reality of the spectrum of one of the {\em other}
observables $\Lambda=\Lambda(\alpha)$. Thus, although the energies
$E_n$ themselves remain, by construction, manifestly insensitive to
any changes of $\alpha$, the eigenvalues of $\Lambda$ will depend on
$\alpha$ in general. In this sense, we may invert the arrow of
argument and reinterpret Eq.~(\ref{Dieud}) as an implicit definition
of the ``admissible" metric for a particular observable
$\Lambda=\Lambda(\alpha)$. Then, naturally, with the change of
$\alpha$ we may reach the {\em physical horizon} (called also an
exceptional point in parametric space \cite{Kato}) at which the
one-parametric spectrum $\{\lambda_j(\alpha)\}_{j=0}^{\infty}$
ceases to be real.

As long as the spectrum of our toy-model Hamiltonian itself remains
real at any $\alpha$ (i.e., formally, its own physical horizon lies
at $\alpha = \pm \infty$), we might call the other, hidden horizons
(i.e., those caused by the boundaries of observability of the other,
not necessarily explicitly, or completely known, observables
$\Lambda$) ``secondary".

\section{Conclusions}

The proposal of consistency of working with quantum observables
which fail to be self-adjoint in a manifestly unphysical, auxiliary
Hilbert space ${\cal K}^{\cal Q}$ dates back to the
nuclear physics inspired analysis in \cite{Geyer}. This approach has
recently been reformulated, in Ref.~\cite{SIGMA}, as a fairly
universal recipe working with a {\em triplet} of Hilbert spaces.
One of these spaces (denoted, in \cite{SIGMA}, as ${\cal H}^{(P)}$
and being ``miscroscopic" or ``fermionic" in the
nuclear physics context of Ref. \cite{Geyer}) may be characterized
as ``physical" but ``prohibitively complicated". In contrast, the
other two (both ``bosonic" in \cite{Geyer}) formed the pair of the
``first auxiliary" space ${\cal H}^{(F)}$ and the ``second
auxiliary" space ${\cal H}^{(S)}$.

The main distinctive feature of this three-Hilbert-space pattern may
be seen in a complete absence of the concept of ${\cal PT}-$symmetry
of the given (and, in the Dirac-metric space ${\cal H}^{(F)}$,
manifestly non-Hermitian) Hamiltonian $H$. Still, in a historical
perspective, the addition of the requirement of the ${\cal
PT}-$symmetry  and of the postulate of the observability of the
charge ${\cal C}^{(PT)}$ proved remarkably productive.

In the original PTQT formalism using the genuine, indefinite parity
${\cal P}$ one starts from the auxiliary Krein space ${\cal K}^{\cal
Q}$ and constructs the Hamiltonian-dependent charge ${\cal C}$ and
the metric (\ref{bend}). The gains (e.g., the productivity and
flexibility of the recipe) are accompanied by certain losses
(e.g., the unexpected emergence of the long-ranged
causality-violating effects caused by the charge ${\cal C} ={\cal
C}(H)$ in the scattering dynamical regime~\cite{Jones}).

Similar limitations may be expected to occur when one turns
attention to the positive definite metrics ${\cal Q}$. A partial
encouragement of such a move may be sought in the recent success of
a {\em simultaneous} introduction of the short-range non-localities
in {\em both} the (genuinely ${\cal PT}-$symmetric) Hamiltonians
$H\neq H^\dagger$ {\em and} band-matrix metrics $\Theta^{(PT)}$
\cite{fund}. In the ``standard", parity-related ${\cal
PT}-$symmetric theory, several families of certain
finite-dimensional Hamiltonians $H$ were already assigned
short-ranged metrics $\Theta$ {\em constructively} \cite{predece}.

All of these PTQT developments sound encouraging and one might
expect that parallel developments could be also realized
in the present QTQT context; we think that our present illustrative
example may also find multiple descendants. One
should emphasize that irrespectively of the indefiniteness or
positivity of ${\cal P}$ resp. ${\cal Q}$, the underlying paradigm
of the Hamiltonian-controlled choice of Hilbert space seems to be
fairly efficient.

After all, as we explained, the recipe is not incompatible even with
the most traditional quantum-theory textbooks. Indeed, in all of the
current implementations of the PTQT/QTQT idea, the apparent
non-Hermiticity $H\neq H^\dagger$ of the ``input" Hamiltonian seems
to be an almost irrelevant, half-hidden byproduct of our habitual
and {\em tacit} preference of the traditional and friendly Hilbert
space ${\cal H}^{(F)}:=L^2(\mathbb{R})$ of square-integrable
functions. In our present proposal, the preference of the Hilbert
spaces in their ``simplest possible" representations ${\cal K}^{\cal
Q}$ has just been given an innovative reinterpretation, with the
simplified initial metric ${\cal Q}$ entering, via
Eq.~(\ref{source}), the closed-form definition of the biorthogonal
basis. Subsequently, such knowledge of the basis converted
formula~(\ref{defini}) into an {\em explicit definition} of all of
the eligible metrics and physical Hilbert spaces ${\cal H}^{(QT)}$.

To summarize, what we have proposed is, in essence, a
double Hilbert space construction of a quantum system from its given
Hamiltonian $H$. Our proposal was inspired by the success of the
methodically productive concept of ${\cal PT}$-symmetry in quantum
mechanics.  With the applicability illustrated via an exactly
solvable Legendre quantum lattice oscillator, our QTQT proposal
refers to ideas and assumptions which are quite similar to their
PTQT predecessors.

The main innovation may be seen in the replacement of the auxiliary,
unphysical Krein space  ${\cal K}^{\cal P}$ (in which the indefinite
operator ${\cal P}$ served as its pseudo-metric) by an equally
unphysical Hilbert space ${\cal K}^{\cal Q}$. For this purpose we
required the positive definiteness of the parity-replacing auxiliary
metric ${\cal Q}$. Still, we tried to keep as many analogies between
${\cal P}$ and ${\cal Q}$ as possible. In particular, we insisted on
the evolution of the system being rendered unitary via a factorized
metric $\Theta^{(QT)}={\cal QC}^{(QT)}$. This being said, we partly
accepted the philosophy of review \cite{ali} and deviated from the
traditional PTQT prescriptions by not trying to make such a metric
unique. This new flexibility enabled us to preserve the
nontriviality of the charge-like operator ${\cal C}^{(QT)}\neq I$,
albeit accompanied by a weakening of the constraints imposed upon
this particular operator.


\vspace{5mm}

\section*{Acknowledgement}

MZ (partially supported by GA\v{C}R grant Nr. P203/11/1433 and by
M\v{S}MT ``Doppler Institute" project Nr. LC06002) acknowledges the
atmosphere and hospitality of STIAS, Stellenbosch where this work
has been mainly performed.


\end{document}